\documentclass[aps,showpacs,nofootinbib,prd,twocolumn]{revtex4-1}
\pdfoutput=1
\usepackage{graphicx}
\usepackage[retainorgcmds]{IEEEtrantools}
\usepackage{amssymb}
\usepackage{amsmath}
\usepackage[svgnames]{xcolor}
\usepackage{mathtools,slashed}
\usepackage{epstopdf}
\usepackage[utf8]{inputenc}
\usepackage{url}
\usepackage[colorlinks,citecolor=DarkGreen,linkcolor=DarkRed,urlcolor=DarkBlue]{hyperref}
\usepackage{subfiles}
\usepackage{physics}
\usepackage[compat=1.1.0]{tikz-feynhand}
\usepackage{tikz}
\usetikzlibrary{arrows.meta}
\usepackage{newtxtext}
\usepackage{newtxmath}
\usepackage[normalem]{ulem}

\usepackage{epsfig}

\newcommand{\be}{\begin{eqnarray}}
\newcommand{\ee}{\end{eqnarray}}
\begin{document}

\title{Relaxation time for the alignment between quark spin and \\ angular velocity  in a rotating QCD medium}

\author{Alejandro Ayala$^1$}
\author{Santiago Bernal-Langarica$^1$}
\author{Isabel Dom\'inguez Jim\'enez$^2$}
\author{Ivonne Maldonado$^{3}$}
\author{José Jorge Medina-Serna$^1$}
\author{Javier Rendón$^1$}
\author{Mar\'ia Elena Tejeda-Yeomans$^{4}$}

  \address{
  $^1$Instituto de Ciencias
  Nucleares, Universidad Nacional Aut\'onoma de M\'exico, Apartado
  Postal 70-543, CdMx 04510,
  Mexico.\\
  %$^2$Centre for Theoretical and Mathematical Physics, and Department of Physics,
  %University of Cape Town, Rondebosch 7700, South Africa.\\
  $^{2}$Facultad de Ciencias Físico-Matemáticas, Universidad Autánoma de Sinaloa, Avenida de las Am\'ericas y Boulevard Universitarios, Ciudad Universitaria, C.P. 80000, Culiac\'an, Sinaloa, M\'exico. \\  
   $^{3}$Joint Institute for Nuclear Research, Dubna 141980, Russia.\\
  $^{4}$Facultad de Ciencias - CUICBAS, Universidad de Colima, Bernal Díaz del Castillo No. 340, Col. Villas San Sebastián, 28045 Colima, Mexico.
}

\begin{abstract}

We compute the relaxation times for massive quarks and antiquarks to align their spins with the angular velocity in a rigidly rotating medium at finite temperature and baryon density. The rotation effects are implemented using a fermion propagator immersed in a cylindrical rotating environment. The relaxation time is computed as the inverse of the interaction rate to produce an asymmetry between the quark (antiquark) spin components along and opposite to the angular velocity. For conditions resembling heavy-ion collisions, the relaxation times for quarks are smaller than for antiquarks. For semicentral collisions the relaxation time is within the possible life-time of the QGP for all collision energies. However, for antiquarks this happens only for collision energies $\sqrt{s_{NN}}\gtrsim 50$ GeV. The results are quantified in terms of the intrinsic quark and antiquark polarizations, namely, the probability to build the spin asymmetry as a function of time. Our results show that these intrinsic polarizations tend to 1 with time at different rates given by the relaxation times with quarks reaching a sizable asymmetry at a faster pace. These are key results to further elucidate the mechanisms of hyperon polarization in relativistic heavy-ion
collisions.
\end{abstract}
\maketitle

\section{Introduction}

Relativistic heavy-ion collisions are the best tool to explore, in a controlled manner, the properties of strongly interacting matter under extreme conditions. The study of the different observables emerging from these reactions has produced a wealth of results revealing an ever more complete picture of these properties for temperatures and densities close to or above the deconfinement transition. However, some other phenomena still miss a clearer understanding and pose a challenge for the evolving standard model of heavy-ion reactions. One of these observables is the relatively large degree of polarization of $\Lambda$ and $\overline{\Lambda}$ hyperons measured in semicentral collision for energies 2.5 GeV $\lesssim \sqrt{s_{NN}}\lesssim 27$ GeV, which shows an increasing trend as the energy and centrality of the collision decreases. The raising trend is different for $\Lambda$s than $\overline{\Lambda}$s~\cite{STAR:2017ckg,STAR:2021beb,STAR:2023nvo,HADES:2022enx}. For semicentral collisions, the matter density profile in the transverse plane induces the development of a global angular momentum, quantified in terms of the thermal vorticity~\cite{Becattini2008,Becattini:2016gvu}. Such angular momentum could be transferred to spin degrees of freedom and be responsible for the observed global polarization~\cite{Becattini:2022zvf}. This expectation is supported by the relation between rotation and spin, nowadays referred to as the Barnett effect, whereby a spinning ferromagnet experiences a change of its magnetization~\cite{1915PhRv....6..239B} and the closely related Einstein–de Haas effect, based on the observation that a change in the magnetic moment of a free body causes this body to rotate~\cite{1915KNAB...18..696E}. As a consequence, significant efforts have been devoted to quantify how this vorticity may be responsible for the magnitude of the observed polarization, assuming that the medium rotation is transferred to the spin polarization regardless of the microscopic mechanisms responsible for the effect~\cite{Karpenko:2021wdm,DelZanna:2013eua,Karpenko:2013wva,Karpenko:2016jyx,Ivanov:2019ern,Wei:2018zfb,Vitiuk:2019rfv,Xie:2019jun,Ivanov:2020udj}. However, the transferring of rotational motion to spin can only happen provided the medium induced reactions occur fast enough so that the alignment of the spin and angular velocity takes place on average within the lifetime of the medium. In the recent literature, this question has been addressed using different approaches~\cite{Montenegro:2018bcf,Kapusta:2019ktm,Kapusta:2019sad,Montenegro:2020paq,Kapusta:2020dco,Kapusta:2020npk,Torrieri:2022ogj}.

In a couple of recent works, we have explored whether this {\it relaxation time} for the alignment is short enough so that the observed polarization of hyperons can be attributed to the transferring of rotation to spin degrees of freedom~\cite{Ayala:2020ndx,Ayala:2019iin}. This is achieved by computing the interaction rate for the spin of a strange quark to align with the thermal vorticity, assuming an effective spin-vorticity coupling in a thermal QCD medium. The findings have been used to compute the $\Lambda$ and $\overline{\Lambda}$ polarization in the context of a core-corona model~\cite{Ayala:2023xyn,Ayala:2022yyx,Ayala:2021xrn,Ayala:2020soy}. The calculation resorts to computing the imaginary part of the self-energy of a vacuum quark whose propagator does not experience the effects of the rotational motion. To improve the description, also in a recent work, we have computed the propagation of a spin one-half fermion immersed in a rigid, cylindrical rotating environment~\cite{Ayala:2021osy}. For these purposes, we have followed the method introduced in Ref.~\cite{Iablokov:2020upc} which requires knowledge of the explicit solutions of
the Dirac equation. These have been previously studied in different contexts by imposing different boundary conditions~\cite{Chodos:1974je,Chernodub:2016kxh,Ambrus:2015lfr,Chen:2015hfc,Ebihara:2016fwa,Ambrus:2014uqa,Yamamoto:2013zwa,Gaspar:2023nqk}.

In this work we use the propagator found in Ref.~\cite{Ayala:2021osy} to compute the imaginary part of the self-energy of a quark immersed in a rotating QCD medium at finite temperature ($T$) and baryo-chemical potential ($\mu_B$). We show that for values of $T$ and $\mu_B$ where the chiral symmetry restoration/deconfinement transition is thought to take place, the relaxation time for quarks turns out to be small enough, compared to the medium life-time, for the  inferred, commonly accepted values of the medium angular velocity, after a semicentral heavy-ion collision. However, this is not the case for the antiquarks except for collision energies $\sqrt{s_{NN}}\gtrsim 50$ GeV. The work is organized as follows: In Sec.~\ref{II} we briefly revisit the derivation of the fermion propagator in a rotating environment. In Sec.~\ref{III} we use this propagator to compute the interaction rate for a quark spin to align with the vorticity in a QCD rotating medium at finite temperature and baryo-chemical potential. In Sec.~\ref{IV} we compute the relaxation time for values of $T$ and $\mu_B$ close to the chiral symmetry restoration/deconfinement transition and show that for quarks this relaxation time is within the putative life-time of the system produced in the reaction, although this is not the case for antiquarks except for large collision energies. We finally summarize and conclude in Sec.~\ref{concl}. 

\section{Propagator for a spin one-half fermion in a rotating environment}\label{II}

The physics within a relativistic rotating frame is most
easily described in terms of a metric tensor resembling
that of a curved space-time. We consider that the interaction region can be thought of as a rigid cylinder rotating around the $\hat{z}$-axis with constant angular velocity $\Omega$  which is produced in semicentral  collisions. 
We can thus write the metric tensor as
\begin{equation}
  g_{\mu\nu}=  \begin{pmatrix}
1-(x^2+y^2)\Omega^2 & y\Omega & -x\Omega & 0\\
y\Omega & -1 & 0& 0\\
-x\Omega & 0&-1 &  0\\
0 & 0 & 0& -1\\
\end{pmatrix}.
\end{equation}
A fermion with mass $m$ within the cylinder is described by
the Dirac equation
\begin{equation}
    \left[i\gamma^\mu\left(\partial_\mu + \Gamma_\mu \right) -m \right]\Psi=0,
    \label{DiracEq}
\end{equation}
where $\Gamma_\mu$ is the affine connection. In this context, the $\gamma^\mu$-matrices in Eq.~(\ref{DiracEq}) correspond to the Dirac matrices in the rotating frame, which satisfy the usual anti-commutation relations 
\begin{equation}
    \{ \gamma^\mu , \gamma^\nu\}=2g^{\mu\nu}.
\end{equation}
The relation between the gamma matrices in the rotating frame and the usual gamma matrices are

\begin{equation}
    \begin{aligned}
        &\gamma^t=\gamma^0, \;\;\;\;\;\;\; \gamma^x=\gamma^1+y\Omega\gamma^0,\\
        &\gamma^z=\gamma^3, \;\;\;\;\;\;\; \gamma^y=\gamma^2-x\Omega\gamma^0.
    \end{aligned}
\end{equation}
In this notation, $\mu=\{t,x,y,z\}$ refers to the rotating frame while $\mu=\{0,1,2,3\}$ refers to the local rest frame. Therefore, Eq.~(\ref{DiracEq}) can be written as
\begin{equation}
\begin{aligned}
      \Big[i\gamma^0 & \left( \partial_t-x\Omega\partial_y + y\Omega\partial_x-\frac{i}{2}\Omega\sigma^{12} \right)\\ &
      +i\gamma^1\partial_x+i\gamma^2\partial_y+i\gamma^3\partial_z -m \Big]\Psi=0. 
\end{aligned}
\label{DiracEq2}
\end{equation}
In the Dirac representation, 
\begin{equation}
  \sigma^{12}=  \begin{pmatrix}
\sigma^3 & 0\\
0 & \sigma^3
\end{pmatrix},
\end{equation}
where $\sigma^3=\mbox{diag}(1,-1)$ is the Pauli matrix associated with the third component of the spin. Therefore, we can rewrite Eq.~(\ref{DiracEq2}) as

\begin{equation}
\begin{aligned}
    \left[i\gamma^0\left(\partial_t+ \Omega\hat{J}_z \right) +i\Vec{\gamma}\cdot\Vec{\nabla}-m \right]\Psi=0,
\end{aligned}
\label{DiracEq3}
\end{equation}
where
\begin{equation}
    \hat{J}_z\equiv\hat{L}_z+\hat{S}_z=-i(x\partial_y - y\partial_x)+\frac{1}{2}\sigma^{12}.
\end{equation}
This expression defines the total angular momentum in the $\hat{z}$ direction. The term $\hat{L}_z$ represents the orbital angular momentum, whereas $\hat{S}_z$ is the spin. On the other hand, the term  $-i\Vec{\nabla}$ is the usual momentum operator. We can find solutions to Eq.~(\ref{DiracEq3}) in the form
\begin{equation}
    \Psi(x)= \left[i\gamma^0\left(\partial_t+ \Omega\hat{J}_z \right) +i\Vec{\gamma}\cdot\Vec{\nabla}+m \right]\phi(x),
    \label{psi}
\end{equation}
and then, the function $\phi(x)$ satisfies a Klein-Gordon like equation
\begin{equation}
    \left[\left(i\partial_t+\Omega\hat{J}_z \right)^2 +\partial^2_x+\partial^2_y+\partial^2_z -m^2\right]\phi(x)=0.
    \label{KG}
\end{equation}
Notice that the spin operator $\hat{S}_z$ when applied to $\phi(x)$ produces eigenvalues $s=\pm$1/2. Consequently, conservation of the total angular momentum expressed in terms of the eigenvalues $j=s+l$ imposes solutions with $l$ for $s=$1/2 and $l+1$ for $s=-$1/2. With these considerations, the solution of Eq.~(\ref{KG}) can be written in cylindrical coordinates $(t, x, y, z) \to (t, \rho \sin\varphi, \rho \cos\varphi, z)$ as
\begin{equation}
    \phi(x)=\begin{pmatrix}
J_l(k_\perp\rho) \\
J_{l+1}(k_\perp\rho)e^{i\varphi} \\
J_l(k_\perp\rho)\\
J_{l+1}(k_\perp\rho)e^{i\varphi}
\end{pmatrix}
e^{-Et+ik_zz+il\varphi},
\end{equation}
where $J_l$ are Bessel functions of the first kind,
\begin{eqnarray}
k_\perp^2=\tilde{E}^2-k_z^2-m^2,
\label{onmassshell}
\end{eqnarray}
is the transverse momentum squared and we have defined $\tilde{E}\equiv E+j\: \Omega$, representing the fermion energy observed from the inertial frame. 
Hence, the solution of Eq.~(\ref{psi}) is
\begin{equation}
\begin{aligned}
        \Psi(x)=&\begin{pmatrix}
\left[E+j\Omega+m-k_z+ik_\perp \right]J_l(k_\perp\rho) \\
\left[E+j\Omega+m+k_z-ik_\perp \right]J_{l+1}(k_\perp\rho)e^{i\varphi} \\
\left[-E-j\Omega+m-k_z+ik_\perp \right]J_l(k_\perp\rho)\\
\left[-E-j\Omega+m+k_z-ik_\perp \right]J_{l+1}(k_\perp\rho)e^{i\varphi}
\end{pmatrix}\\
&\times e^{-(E+j\Omega)t+ik_zz+il\varphi}.
\end{aligned}
\end{equation}
Before writing the expression for the fermion propagator in the rotating environment, it is important to highlight some features of the solution. First, causality requires that $\Omega R<1$, where $R$ is the radius of the cylinder. Therefore, the solution is valid as long as $\Omega < 1/R$. Second, we can simplify the solution assuming that the fermion is totally dragged by the vortical motion such that the angular position is determined by the product of the angular velocity and the time, specifically $\varphi +\Omega t=0$. This is a reasonable approximation when  considering that during the early stages of a peripheral heavy-ion collision, particle interactions have not yet produced the development of a radial expansion. With this approximation, the propagator is translational invariant and can be simply Fourier transformed. With these features in mind, we write the fermion propagator $S(x,x')$ as
\begin{widetext}
\begin{equation}
   S(x,x') =\left[i\gamma^0\left(\partial_t+ \Omega\hat{J}_z \right) +i\Vec{\gamma}\cdot\Vec{\nabla}+m \right]G(x,x'),
   \label{Prop1}
\end{equation}
where
\begin{equation}
    G(x,x')=(-i)\int_{-\infty}^0 d\tau\sum_\lambda\exp{\left[-i\tau\lambda \right]}\phi_\lambda(x)\phi_\lambda^\dag(x).
\end{equation}
In this last expression, $\lambda$ and $\phi_ \lambda (x)$ represent the eigenvalues and eigenvectors of Eq.~(\ref{KG}). Taking $E,k_\perp,k_z,l$ as independent quantum numbers, the closure relation is written as
\begin{equation}
      \sum_\lambda\phi_\lambda(x)\phi_\lambda^\dag(x) =\sum_{l=\infty}^\infty\int\frac{dEdk_zdk_\perp k_\perp}{(2\pi)^3}\phi(x)\phi^\dag(x)\\ 
=\begin{pmatrix}
1&0&1&0\\
0&1&0&1\\
1&0&1&0\\
0&1&0&1
\end{pmatrix}\delta^4(x-x').
\end{equation}
Hence, Eq.~(\ref{Prop1}) becomes
\begin{equation}
    S(x,x') =\sum_{l=\infty}^\infty\int\frac{dEdk_zdk_\perp k_\perp}{(2\pi)^3}\Phi(\rho,\rho')\frac{e^{-i(E-(l+1/2)\Omega)(t-t')}e^{ik_z(z-z')}e^{il(\varphi-\varphi')}}
    {E^2-k_z^2-m^2-k_\perp^2+i\epsilon},
 \label{Prop2}
\end{equation}
where
\begin{equation}
   \Phi(\rho,\rho')=\begin{pmatrix}
 \mathcal{A}\mathcal{J}_{l,l} & \mathcal{A}\mathcal{J}_{l,l+1}e^{-i\varphi'} & \mathcal{A}\mathcal{J}_{l,l} & \mathcal{A}\mathcal{J}_{l,l} \\
  \mathcal{B}\mathcal{J}_{l,l+1}e^{i\varphi} & \mathcal{B}\mathcal{J}_{l+1,l+1}e^{i(\varphi-\varphi')} & \mathcal{B}\mathcal{J}_{l+1,l}e^{i\varphi} & \mathcal{B}\mathcal{J}_{l+1,l+1}e^{i(\varphi-\varphi')} \\
  \mathcal{C}\mathcal{J}_{l,l} & \mathcal{C}\mathcal{J}_{l,l+1}e^{-i\varphi'} & \mathcal{C}\mathcal{J}_{l,l} & \mathcal{C}\mathcal{J}_{l,l} \\
    \mathcal{D}\mathcal{J}_{l,l+1}e^{i\varphi} & \mathcal{D}\mathcal{J}_{l+1,l+1}e^{i(\varphi-\varphi')} & \mathcal{D}\mathcal{J}_{l+1,l}e^{i\varphi} & \mathcal{D}\mathcal{J}_{l+1,l+1}e^{i(\varphi-\varphi')} 
\end{pmatrix},
\end{equation}
and we have defined
\begin{equation}
    \begin{aligned}
        \mathcal{A}&\equiv\left[ E+m-k_z+ik_\perp \right],\\
        \mathcal{B}&\equiv\left[ E+m+k_z-ik_\perp \right],\\
        \mathcal{C}&\equiv\left[ -E+m-k_zk+ik_\perp \right],\\
        \mathcal{D}&\equiv\left[ -E+m+k_zk-ik_\perp \right],\\
    \end{aligned}
\end{equation}
with
\begin{equation}
    \mathcal{J}_{l,l'}\equiv J_l(k_\perp\rho)J_{l'}(k_\perp\rho').
\end{equation}
Carrying out the integration and the summation, and taking the Fourier transform, we obtain
\begin{equation}
    S(p)=\begin{pmatrix}
\frac{p_0+\Omega/2-p_z+m+ip_\perp}{\left( p_0+\Omega/2 \right)^2-p^2-m^2+i\epsilon} & 0 &\frac{p_0+\Omega/2-p_z+m+ip_\perp}{\left( p_0+\Omega/2 \right)^2-p^2-m^2+i\epsilon}&0\\
0 & \frac{p_0+\Omega/2+p_z+m-ip_\perp}{\left( p_0-\Omega/2 \right)^2-p^2-m^2+i\epsilon} & 0 & \frac{p_0+\Omega/2+p_z+m-ip_\perp}{\left( p_0-\Omega/2 \right)^2-p^2-m^2+i\epsilon}\\
\frac{-(p_0+\Omega/2)-p_z+m+ip_\perp}{\left( p_0+\Omega/2 \right)^2-p^2-m^2+i\epsilon} & 0 &\frac{-(p_0+\Omega/2)-p_z+m+ip_\perp}{\left( p_0+\Omega/2 \right)^2-p^2-m^2+i\epsilon}&0\\
0 & \frac{-(p_0+\Omega/2)+p_z+m-ip_\perp}{\left( p_0-\Omega/2 \right)^2-p^2-m^2+i\epsilon} & 0 & \frac{-(p_0+\Omega/2)+p_z+m-ip_\perp}{\left( p_0-\Omega/2 \right)^2-p^2-m^2+i\epsilon}
\end{pmatrix}.
\label{Prop2}
\end{equation}  
We can write Eq.~(\ref{Prop2}) in terms of the Dirac-gamma matrices as
 \begin{equation}
        S(P)=\frac{\left(p_0+\Omega/2-p_z+ip_{\perp}\right)\left(\gamma_0+\gamma_3\right)+m\left(1+\gamma_5\right)}{(p_0+\Omega/2)^2-p^2-m^2+i\epsilon}\mathcal{O}^+
            +\frac{(p_0-\Omega/2+p_z-ip_{\perp})(\gamma_0-\gamma_3)+m(1+\gamma_5)}{(p_0-\Omega/2)^2-p^2-m^2+i\epsilon}\mathcal{O}^-,
    \label{PropRot}
    \end{equation}
\end{widetext}
where
\begin{equation}
    \mathcal{O}^{\pm}=\frac12\left[1\pm i\gamma^1\gamma^2\right]
\end{equation}
is the spin projection operator. Notice that the derivation of the fermion propagator is performed in vacuum. To use this propagator including a finite temperature and chemical potential, recall that in equilibrium it is sufficiently general to make the replacement $p_0\to i\tilde{\omega}_n+\mu$ where $\omega_n=(2n+1)\pi T$ are Matsubara frequencies for fermions. Equation~(\ref{PropRot}) represents our approximation for the fermion propagator in a cylindrical rigidly rotating environment. We now use this propagator to compute the relaxation time for the fermion spin to align with the angular velocity in the rotating medium.

\section{Interaction rate for a quark spin to align with the angular velocity in a QCD rotating medium}\label{III}

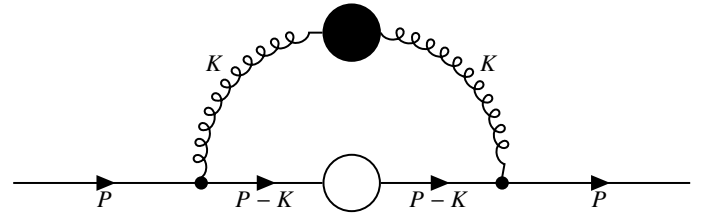
\begin{figure}[b]
    \centering
    \begin{tikzpicture}
			\begin{feynhand}
				\vertex (a) at (0,0); \vertex [dot] (b) at (2.5,0) {}; \vertex [blob] (c) at (4.5,2) {}; \vertex [dot] (d) at (6.5,0) {}; \vertex (e) at (9,0); \vertex [ringblob] (f) at (4.5,0) {};
				\propag [fer] (a) to [edge label' =$P$] (b);
				\propag [fer] (b) to [edge label' =$P-K$] (f);
				\propag [fer] (f) to [edge label' = $P-K$] (d);
				\propag [fer] (d) to [edge label' =$P$] (e) ;
				\propag [glu] (b) to [quarter left, edge label =$K$] (c);
				\propag [glu] (c) to [quarter left, edge label=$K$] (d);
			\end{feynhand}
	\end{tikzpicture}
    \caption{One-loop quark self-energy diagram that defines the
kinematics. The gluon line with a blob represents the effective
gluon propagator at finite density and temperature. The open circle on
the fermion propagator represents the effect of the rotating environment.}
    \label{loop}
\end{figure}

In a QCD plasma in thermal equilibrium at temperature $T$ and baryon chemical potential $\mu_B$, the interaction rates $\Gamma^\pm$ for a quark with spin components $s=\pm 1/2$ in the direction of $\vec{\Omega}$ and four-momentum $P=(p_0,\Vec{p})$ to align its spin in the direction of the angular momentum vector can be expressed in terms of the total interaction rate, which in turn is given by the probability (per unit time) for a transition between the same quantum quark state $u^\pm$, represented by properly normalized spinors with a definite spin projection ($\pm$)  along the direction of the angular velocity. This transition is mediated by the imaginary part of the self-energy, Im$\Sigma$. In symbols,

\begin{equation}
\Gamma^\pm(p_0)\sim \Bar{u}_a^\pm\text{Im}\Sigma_{ab}u_b^\pm.
\label{gammadef1}
\end{equation}
Shuffling the indexes around, we can also write
\begin{eqnarray}
\Gamma^\pm(p_0)&\sim& u_b^\pm\Bar{u}_a^\pm\text{Im}\Sigma_{ab}\nonumber\\
&=&\text{Tr}[\mathcal{O}^\pm\text{Im}\Sigma_{ab}],    
\label{gammadef2}
\end{eqnarray}
where we used that the spin projection operators $\mathcal{O}^\pm$ are given by

\begin{eqnarray}
    \mathcal{O}^\pm\equiv u^\pm\Bar{u}^\pm.
\end{eqnarray}
To extract the {\it creation rate} for e for spin-aligned quark states
from the total interaction rate, as discussed in Ref.~\cite{LeBellac}, we
multiply the total interaction rate by the fermion distribution function $\tilde{f}$, for a grand-canonical ensemble in the presence of a conserved charge to which a quark chemical potential $\mu=1/3 \mu_B$ is associated, namely,

\begin{eqnarray}
    \Gamma^\pm(p_0)= \tilde{f}(p_0 - \mu \mp \Omega / 2)\text{Tr}\left[O^\pm\;\text{Im}\Sigma\right].
\label{gammadef}
\end{eqnarray}
In previous analyses~\cite{Ayala:2020ndx,Ayala:2019iin} the interaction has been modeled using an effective vertex coupling the thermal vorticity and the quark spin. To improve the description, hereby we consider the case where the fermion is subject to the effect of a rotation within a rigid cylinder. The one-loop  contribution to $\Sigma$, depicted in Fig.~\ref{loop}, is given by

\begin{equation}
\Sigma(P)=T\sum_n\int\frac{d^3k}{(2\pi)^3}\gamma^\mu T_a S(P-K)\gamma^\nu T_b \;^*G_{\mu\nu}^{ab},
\end{equation}
where $S$ is the quark propagator in a rotating environment obtained in Eq.~(\ref{PropRot}), $^*G_{\mu\nu}^{ab}$ is the effective gluon propagator in the thermal medium and $T_a$ are the $SU(3)$ group generators. The four-momenta are $P=(i\Tilde{\omega},\vec{p})$ for the fermion and $K=(i\omega_n,\vec{k})$ for the gluon, with $\omega_n$ being the gluon Matsubara frequencies. Also, $^*G_{\mu\nu}^{ab}=^*\!\!G_{\mu\nu}\delta^{ab}$ and
in the hard thermal loop (HTL) approximation $^* G_{\mu\nu}$ is given by
\begin{equation}
    ^*G_{\mu\nu}(K)=\Delta_L(K)P_{L\;\mu\nu} +\Delta_T(K)P_{T\;\mu\nu}
\end{equation}
where $P_{L,T\;\mu\nu}$ are the polarization tensors for three-dimensional longitudinal and transverse gluons~\cite{LeBellac}. The gluon propagator functions for longitudinal and transverse modes, $\Delta_{L,T}(K)$, are given by
\begin{equation}
    \Delta_L^{-1}(K)=K^2+2m_T ^2\frac{K^2}{k^2}\left[1-\frac{i\omega_n}{k}Q_0\left(\frac{i\omega_n}{k}\right)\right],
\end{equation}
\begin{eqnarray}
    \Delta_T^{-1}(K)&=&-K^2-m_T ^2\left(\frac{i\omega_n}{k}\right)\nonumber\\
    &\times&\left\{\left[1-\left(\frac{i\omega_n}{k}\right)^2\right]Q_0\left(\frac{i\omega_n}{k}\right)+\left(\frac{i\omega_n}{k}\right)\right\},
\end{eqnarray}
where 
 \begin{equation}
    Q_0(x)=\frac{1}{2}\ln\left(\frac{x+1}{x-1}\right),
\end{equation}
and $m_T^2$ is the gluon thermal mass squared given by
\begin{equation}
    m_T ^2=\frac{1}{6}g^2\mathcal{C}_AT^2+\frac{1}{12}g^2\mathcal{C}_F(T^2+\frac{3}{\pi^2}\mu^2),
\end{equation}
where $\mathcal{C}_A$ and $\mathcal{C}_F$ are the Casimir factors for
the adjoint and fundamental representations of $SU(3)$.

It is convenient to first look at the sum over Matsubara frequencies for the products of the propagator functions for longitudinal and transverse
gluons, $\Delta_i$ with $i=L,T$, and the Matsubara propagator for the quark in a rotating environment $\tilde{\Delta}$, wich is described in Ref.~\cite{LeBellac}, can be obtained as the inverse of the denominator of each of the components of Eq.~\ref{PropRot} with the replacement  $p_0\to i\tilde{\omega}_n+\mu$.
\begin{equation}
S_i(i\omega)=T\sum_n\Delta_i(i\omega_n)\tilde{\Delta}(i(\omega-\omega_n)).    
\end{equation}
The sum can be performed introducing the spectral densities $\rho_i$ and $\rho_F$ for the gluon and fermion, respectively. The imaginary part of $S_i$ can be written as
\begin{eqnarray}
    \text{Im}(S_i)\!\!\!&=&\!\!\!\pi(e^{\beta(p_0-\mu - \Omega / 2)}+1)\int_{-\infty}^{\infty}\int_{-\infty}^{\infty}\frac{dk_0}{2\pi}\frac{dp'_0}{2\pi}f(k_0)\nonumber\\
    \!\!\!\!\!\!\!&\times&\!\!\tilde{f}(p'_0-\mu \mp \Omega / 2) \delta(p_0-k_0-p'_0)\nonumber \\
    & \times &\rho_i(k_0,k)\rho_F(p'_0,p-k),
\label{Matsu1}
\end{eqnarray}
where $f(k_0)$ is the Bose-Einstein distribution. The spectral densities $\rho_i$ are obtained from the imaginary part of $\Delta_i(i\omega_n)$ after the analytic continuation $i\omega_n\rightarrow k_0+i\epsilon$ and contain the discontinuities of the gluon propagator across the real $k_0$ axis. Their support depends on the ratio $x=k_0/k$. For $|x|>1$, $\rho_i$ have support on the (timelike) quasi-particle poles. For $|x|<1$, their support coincides with the branch cut of $Q_0(x)$ and corresponds to Landau damping. On the other hand, the fermion spectral density is
\begin{equation}
        \rho_F(p_0',p)=-2\pi\delta \left((p_0'\pm\Omega/2)^2-p^2-m^2\right),\\
    \label{rho}
    \end{equation}
We now concentrate on the trace factors required for the computation of Eq.~(\ref{gammadef}). The term proportional to the fermion momentum and angular velocity
\begin{eqnarray}
     P_{L,T\;\mu\nu}\text{Tr}\left[\gamma^\mu\left( \gamma^0\pm\gamma^3\right)\left(1\pm i\gamma^1\gamma^2\right)\gamma^\nu \right]=0,
     \label{TraceMomentum}
\end{eqnarray}
\begin{widetext}
vanishes identically, whereas the terms proportional to the fermion mass are given by
\begin{eqnarray}
P_{L\;\mu\nu}\text{Tr}\left[\gamma^\mu\left(1+\gamma^5\right)\left(1\pm i\gamma^1\gamma^2\right)\gamma^\nu \right]&=&-4\nonumber\\
P_{T\;\mu\nu}\text{Tr}\left[\gamma^\mu\left(1+\gamma^5\right)\left(1\pm i\gamma^1\gamma^2\right)\gamma^\nu \right]&=&-8.
\end{eqnarray}

The delta functions in Eqs.~\eqref{Matsu1} and~\eqref{rho} restrict the integration over gluon energies to the spacelike region, $|x|<1$. Therefore, the parts of the gluon spectral densities that contribute to the interaction rate are given by
\begin{equation}
    \rho_L (k_0,k)  =  \frac{2\pi m_T ^2 x \theta(1-x^2)}{\left[k^2 + 2m_T ^2 \left(1-\frac{x}{2}\ln\left\vert\frac{x+1}{x-1}\right\vert\right)\right]^2 + \pi^2 m_T ^4 x^2},
\end{equation}
\begin{equation}
    \rho_T (k_0,k)  =  \frac{2\pi m_T ^2 x(1-x^2)\theta(1-x^2)}{2\left[\left(k^2(x^2-1) - m_T ^2 \left[x^2 + \frac{x(1-x^2)}{2}\ln\left\vert \frac{x+1}{x-1}\right\vert\right]\right)^2 + \frac{\pi^2}{4} m_T ^4 x^2 (1-x^2)^2 \right]}.
\end{equation}
\end{widetext}

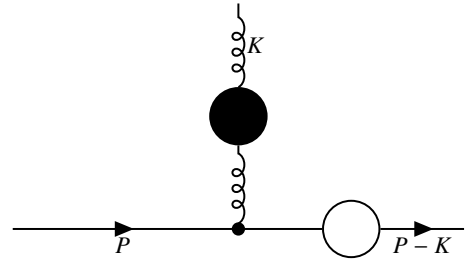
\begin{figure}[b]
    \centering
       \begin{tikzpicture}
			\begin{feynhand}
			     \vertex (a) at (-3,0);
            \vertex [dot] (b) at (0,0) {};
            \vertex [ringblob] (c) at (1.5,0) {};
            \vertex (d) at (3,0);
            \vertex [blob] (e) at (0,1.5) {};
            \vertex (f) at (0,3);
            \propag [fer] (a) to [edge label' = $P$](b);
            \propag [plain] (b) to (c);
            \propag [fer] (c) to [edge label' = $P-K$] (d);
            \propag [glu] (b) to (e);
            \propag [glu] (e) to [edge label' = $K$] (f);
			\end{feynhand}
	\end{tikzpicture}
    \caption{Feynman diagram representing a process whereby an initially nonrotating quark is dragged by the medium and aligns its spin either parallel or antiparallel to the angular velocity by means of its interactions with medium particles mediated by soft thermal gluons.}
    \label{FP2}
\end{figure}

With all these ingredients we write the interaction rate as
\begin{IEEEeqnarray}{rCl}
        \Gamma^\pm (p_0) & = & \frac{g^2m C_F\pi}{2} \int \frac{d^3 k}{(2\pi)^3} \int_{-\infty} ^\infty \frac{dk_0}{2\pi} \int_{-\infty} ^\infty dp_0 ^\prime \nonumber\\
       &\times &
       f(k_0)\left(4\rho_L(k_0) + 8\rho_{T}(k_0)  \right)\nonumber \\
&  \times &\tilde{f}(p_0 ^\prime - \mu \mp \Omega / 2) \delta(p_0 - k_0 - p_0 ^\prime)\nonumber\\  
&\times & \delta\left(\left(p_0 ^\prime \pm \Omega/2\right)^2 - E^2\right),
    \label{gama0}
\end{IEEEeqnarray}
with 
\begin{equation}
    E^2=|\vec{p}-\vec{k}|^2-m^2.
\end{equation}
Notice that
\begin{eqnarray}
    &&\delta\left(\left(p_0' \pm \Omega/2\right)^2 - E^2\right)=\nonumber\\
    &&\frac{1}{2E}\Big[\delta(p_0'\pm\Omega/2 - E)+\delta(p_0^\prime\pm\Omega/2 + E) \Big].
\end{eqnarray}
Therefore, we can integrate Eq.~(\ref{gama0}) over $p_0^\prime$ to obtain
\begin{IEEEeqnarray}{rCl}
        \Gamma^\pm (p_0) & = & \frac{g^2m C_F\pi}{2} \int \frac{d^3 k}{(2\pi)^3} \int_{-\infty} ^\infty \frac{1}{2E}\frac{dk_0}{2\pi} f(k_0) \nonumber\\
       &\times& \left(4\rho_L(k_0) + 8\rho_{T}(k_0)  \right) \nonumber\\
&  \times & \Big[\tilde{f}(E - \mu\mp\Omega)\delta(p_0 - k_0 - E \pm\Omega/2)\nonumber\\
&+ &\tilde{f}(-E - \mu\mp\Omega)\delta(p_0 - k_0 + E \pm\Omega/2)\Big].
    \label{gama1}
\end{IEEEeqnarray}
Notice that for the considered angular velocities appropriate to the early stages of the collision (10 MeV $\lesssim \Omega\lesssim 14$ MeV), and for a strange quark mass $m\sim 100$ MeV, the combination $E\pm \Omega/2$ can always be safely regarded as being positive. The kinematical constraint imposed by the first of the delta functions in Eq.~(\ref{gama1}) corresponds to the rate to produce rotating and thermalized quarks originated by the dispersion of vacuum nonrotating quarks as a result of dispersion with medium quarks. This is depicted in Fig.~\ref{FP2}.
We then single out this contribution from the total rate which can then be written as
\begin{IEEEeqnarray}{rCl}
        \Gamma^\pm (p_0) & = & \frac{g^2m C_F\pi}{2} \int \frac{k^2dk\;d(\cos\theta)d\phi}{(2\pi)^3} \int_{-\infty} ^\infty \frac{1}{2E}\frac{dk_0}{2\pi} \nonumber \\
       &\times & \left(4\rho_L(k_0) + 8\rho_{T}(k_0)  \right)\delta(p_0 - k_0 - E \pm\Omega/2) \nonumber\\
&\times &  f(k_0) \tilde{f}(E -\mu\mp\Omega).
    \label{gama2}
\end{IEEEeqnarray}

The kinematical restrictions for the  $k_0$ integration
translate into integration regions $\mathcal{R^\pm}$. After integrating over the angle $\theta$ between $\vec{p}$ and $\vec{k}$, and over the azimuthal angle $\phi$, and finally using that $E^2=p^2+\ m^2=|\vec{p}-\vec{k}|^2-m^2=p^2+k^2-2pk\cos{\theta}+m^2$, we obtain
\begin{IEEEeqnarray}{rCl}
\Gamma^\pm (p_0)  & = & \frac{g^2m C_F\pi}{2}\int_0 ^\infty \frac{dk\,k^2}{(2\pi)^3}\int_{\mathcal{R^\pm}}dk_0 \frac{f(k_0)}{2pk}\nonumber\\
&\times& \left(4\rho_L(k_0) + 8\rho_{T}(k_0)  \right)\tilde{f}\left(p_0 - k_0 - \mu \mp \Omega/2\right).\IEEEeqnarraynumspace
    \label{GammaT1}
\end{IEEEeqnarray}
where $\mathcal{R^\pm}$ are the regions defined by
\begin{equation}
\begin{aligned}
  k_0 &\leq p_0\pm\Omega/2 - \sqrt{(p-k)^2+m^2},\\
 k_0 &\geq p_0\pm\Omega/2 - \sqrt{(p+k)^2+m^2}.
\end{aligned}
\end{equation}
The total rate to align the quark spin with the angular velocity is thus given by the difference between the rate to populate the spin projection along and opposite to the angular velocity which is obtained by
integrating the difference between $\Gamma^+$ and $\Gamma^-$ of Eq.~(\ref{GammaT1}) over the available phase space
\begin{equation}
    \Gamma=V\int\frac{d^3p}{(2\pi)^3}\left[\Gamma^+(p_0)-\Gamma^-(p_0)\right],
    \label{gammaTot}
\end{equation}
where $V$ is the volume of the collision region.
\begin{figure}[t]
    \centering
    \includegraphics[scale=0.45]{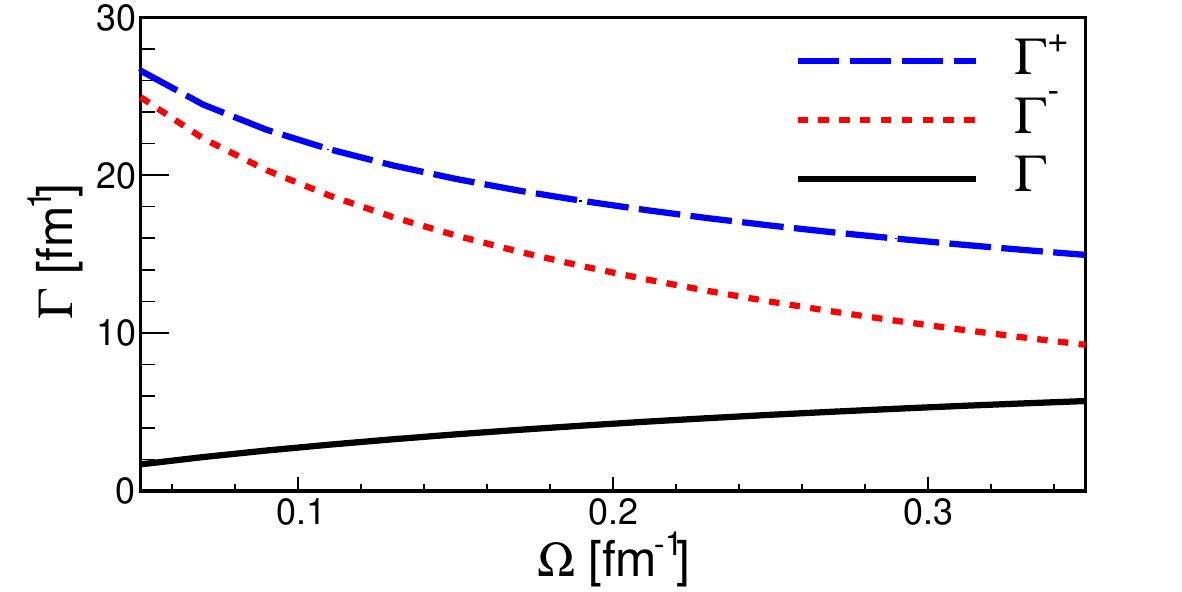}
    \caption{Interaction rates $\Gamma^\pm$ for positive (+) and negative (-) spin projections for quarks as functions of the angular velocity $\Omega$ for semicentral  collisions at an impact parameter $b$ = 10 fm and chemical potential $\mu=$ 100 MeV for a temperature $T$ = 150 MeV. Shown is also the interaction rate $\Gamma$ obtained as the phase space integrated difference $\Gamma^+(p_0) - \Gamma^-(p_0)$.}
    \label{Gamm}
\end{figure}
\begin{figure}[b]
    \centering
    \includegraphics[scale=0.45]{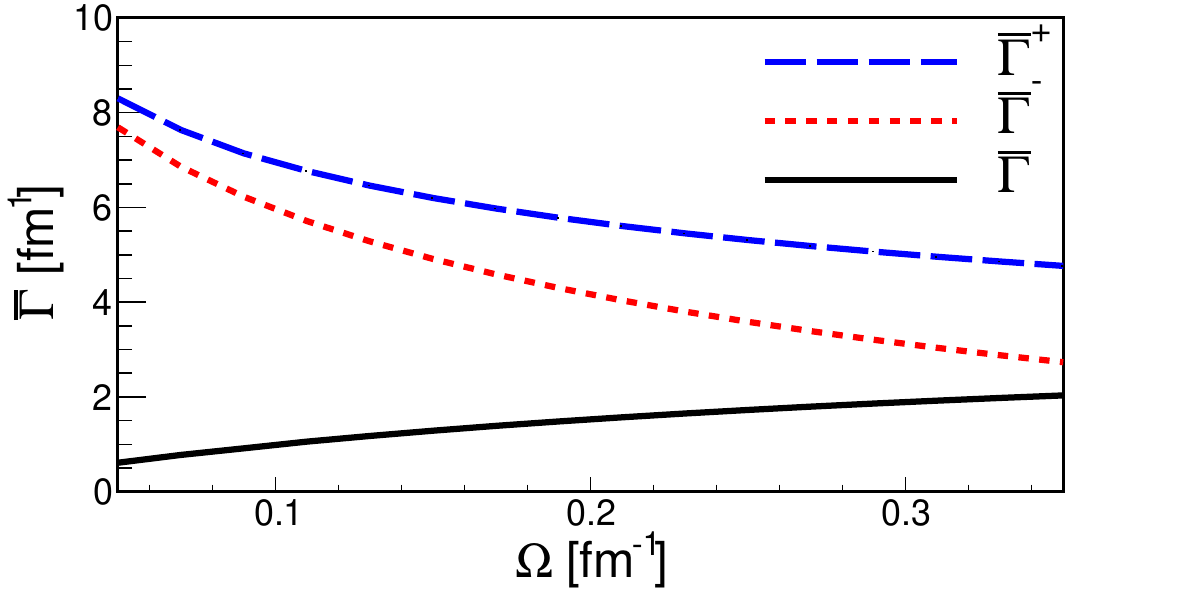}
    \caption{Interaction rates $\Bar{\Gamma}^\pm$ for positive (+) and negative (-) spin projections for antiquarks as functions of the angular velocity $\Omega$ for semicentral  collisions at an impact parameter $b$ = 10 fm and chemical potential $\Bar{\mu}=$ 100 MeV for a temperature $T$ = 150 MeV. Shown is also the interaction rate $\Bar{\Gamma}$ obtained as the phase space integrated difference $\Bar{\Gamma}^+(p_0) - \Bar{\Gamma}^-(p_0)$.}
    \label{Gamm2}
\end{figure}
To compute $V$ for conditions that depend on the
collision energy, we consider a Bjorken expansion scenario
where the volume and the QGP lifetime are related by
\begin{equation}
    V = \pi R^2 \Delta \tau_{QGP},
\end{equation}
where $R$ is the radius of the colliding species and $\Delta\tau_{QGP}$ is the QGP lifetime, which is given as the interval elapsed from the initial formation $\tau_0$ until the hadronization time $\tau_f$~\cite{Ayala:2021xrn}.

Figure~\ref{Gamm} shows the interaction rates $\Gamma^\pm$ for positive and negative quark spin projections, respectively, as functions of the angular velocity $\Omega$, for semicentral collisions at an impact parameter $b$ = 10 fm and chemical potential $\mu=$ 100 MeV for quarks. Since the occupation numbers for antiquarks are obtained from the occupation numbers for quarks by the replacement $\mu\to-\mu\equiv\Bar{\mu}$, the corresponding rates for antiquarks $\Bar{\Gamma}^\pm$ are computed performing such replacement in Eq~\ref{GammaT1}. Figure~\ref{Gamm2} shows the rates for the antiquarks using $\Bar{\mu}=$100 MeV. In both cases, we use a temperature $T=150$ MeV. Notice the symmetry $\Bar{\Gamma}(-\mu)=\Gamma(\mu)$. Also, hereafter, we take the value of the
strong coupling as $\alpha_s=0.3$, and, although the calculation is valid for any quark with nonvanishing mass, we conside the computation of the relaxation time for the case of a strange quark/antiquark mass $m=100$ MeV. This is chosen having in mind to later use the results in the context of the computation of the $\Lambda$ and $\Bar{\Lambda}$ polarizations, under the assumption that the whole hyperon polarizations come from the strange-quark polarization.\\
Shown in the figures are also the phase space integrated differences $\Gamma^+(p_0)-\Gamma^-(p_0)$ and  $\Bar{\Gamma}^+(p_0)-\Bar{\Gamma}^-(p_0)$, respectively, which represent the rates to align the quark or antiquark spin with the angular velocity. Notice that although the rates $\Gamma^\pm$ and $\Bar{\Gamma}^\pm$ are both decreasing functions, their difference increases with $\Omega$. This means that overall the rate at which the positive spin component dominates over the negative one increases as the angular velocity increases. The decrease of the individual rates with $\Omega$ can be traced back to Eq.~\ref{PropRot} that for large $\Omega$ decrease as $1/\Omega$. This behavior is translated to the fermion spectral density
and through it to the region of integration and ultimately to
each of the reaction rates.\\
From the expression for $\Gamma$ in Eq.~(\ref{gammaTot}) we can find the parametric dependence of the relaxation time for spin and angular velocity alignment, defined as
\begin{equation}
    \tau\equiv 1/\Gamma,
\end{equation}
which we proceed to compute.

\section{Relaxation time}\label{IV}

\begin{figure}[t]
    \centering
    \includegraphics[scale=0.45]{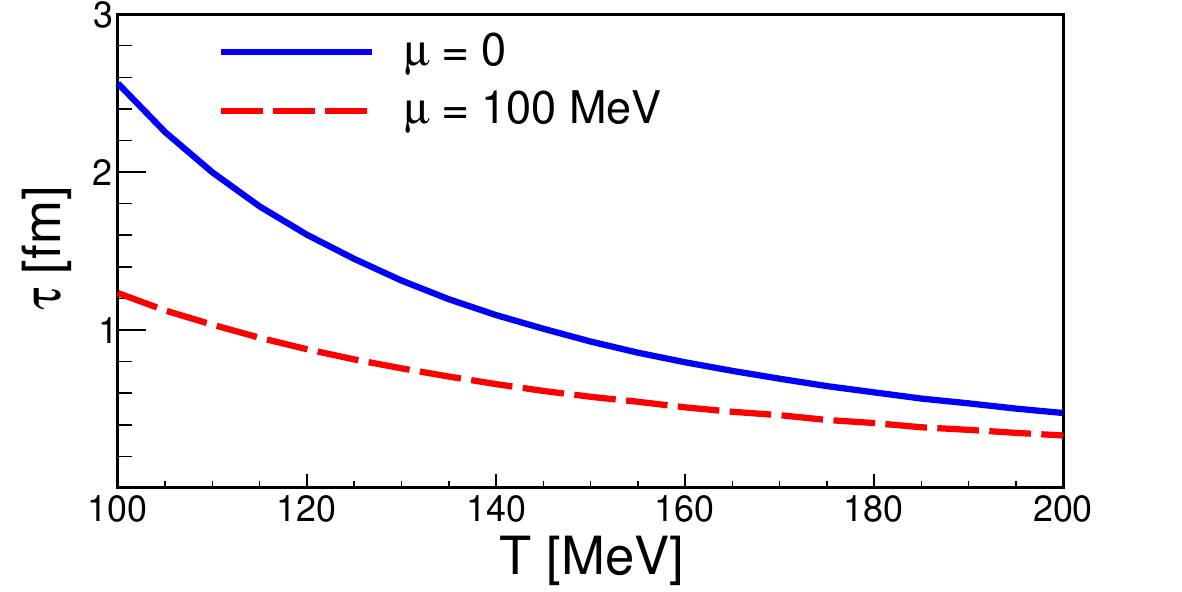}
    \caption{Relaxation time $\tau$ for quarks as a function of the temperature $T$ for semicentral  collisions at an impact parameter $b$ = 10 fm for $\sqrt{s_{NN}}=$ 200 GeV which corresponds to a angular velocity $\Omega=$ 0.052 $\text{fm}^{-1}$ with $\mu=$ 0, 100 MeV.}
    \label{Tau1}
\end{figure}
\begin{figure}[b]
    \centering
    \includegraphics[scale=0.45]{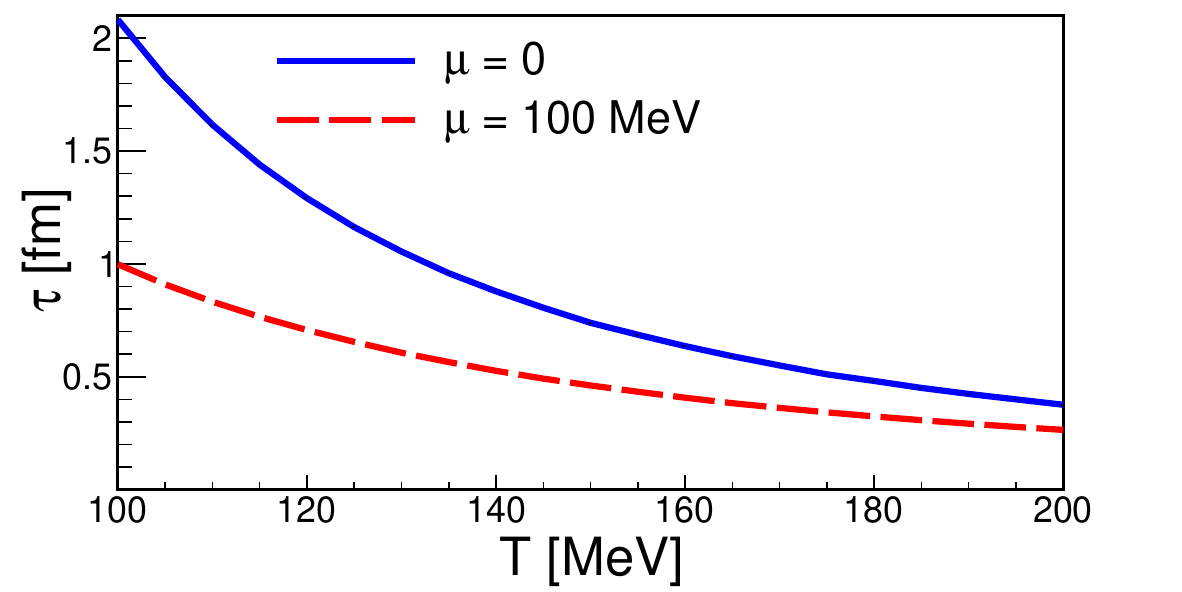}
    \caption{Relaxation time $\tau$ for quarks as a function of the temperature $T$ for semicentral  collisions at an impact parameter $b$=10 fm for $\sqrt{s_{NN}}=$ 10 GeV which corresponds to a angular velocity $\Omega=$0.071 $\text{fm}^{-1}$ with $\mu=$ 0, 100 MeV.}
   \label{Tau2}
\end{figure}

We now concentrate on the computation of the relaxation time when varying the parameters involved in the calculation. For a direct comparison with previous results, we will use the values obtained in \cite{Ayala:2020ndx} for the initial angular velocity. 

\begin{figure}[t]
    \centering
    \includegraphics[scale=0.45]{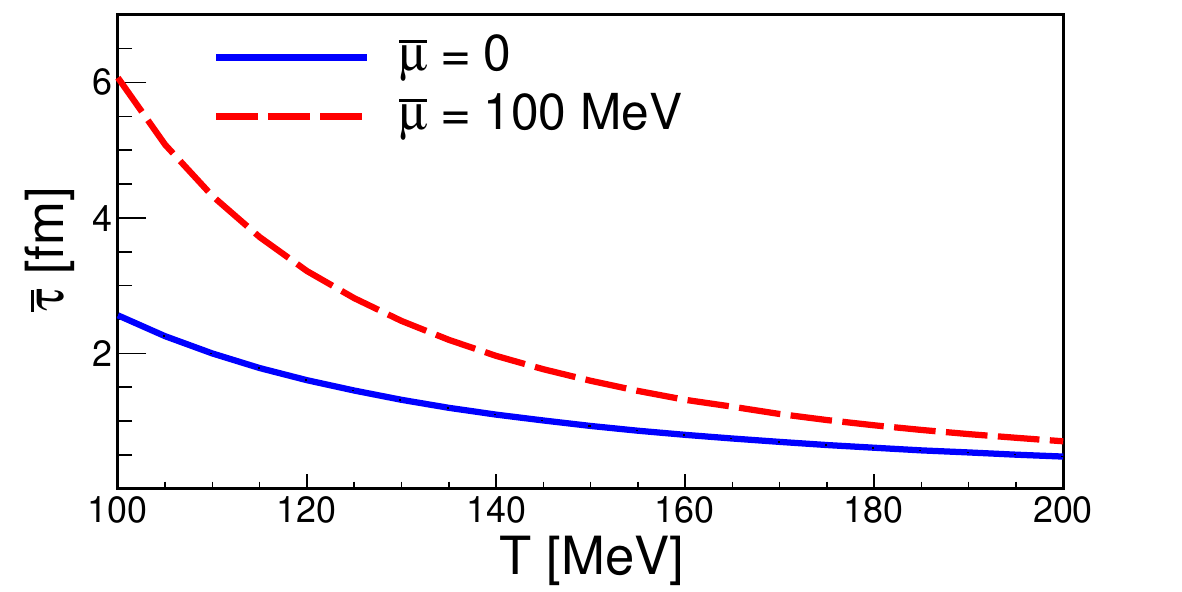}
    \caption{Relaxation time $\Bar{\tau}$ for antiquarks as a function of the temperature $T$ for semicentral  collisions at an impact parameter $b$=10 fm for $\sqrt{s_{NN}}=$ 200 GeV which corresponds to a angular velocity $\Omega=$ 0.052 $\text{fm}^{-1}$ with $\Bar{\mu}=$ 0, 100 MeV.}
    \label{A1}
\end{figure}
\begin{figure}[b]
    \centering
    \includegraphics[scale=0.45]{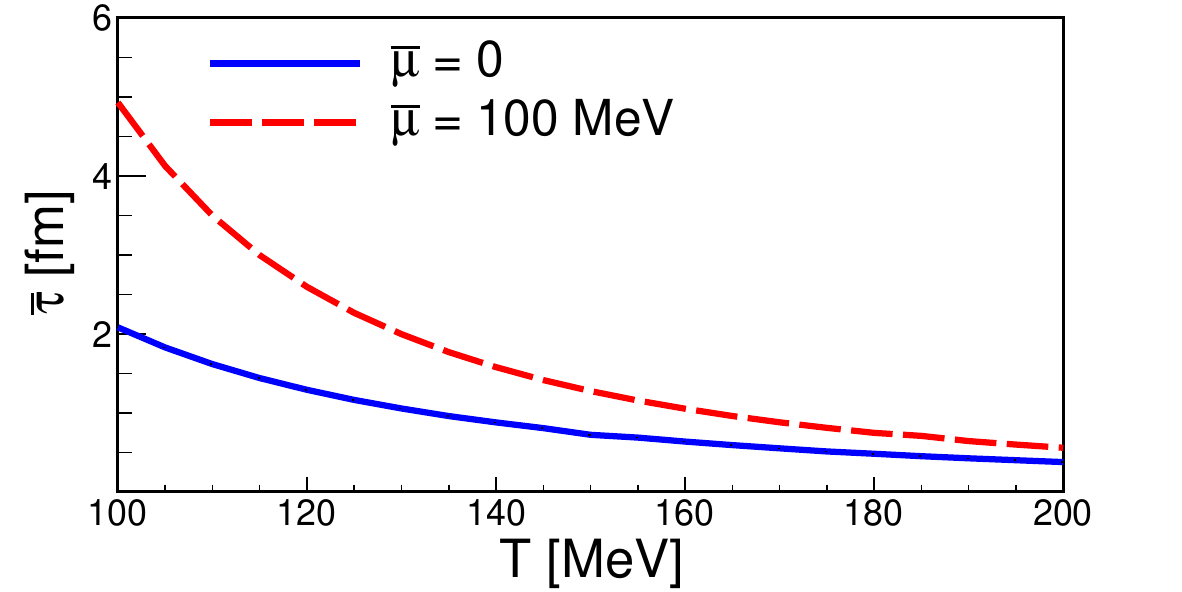}
    \caption{Relaxation time $\Bar{\tau}$ for antiquarks as a function of the temperature $T$ for semicentral  collisions at an impact parameter $b=10$ fm for $\sqrt{s_{NN}}=10$ GeV which corresponds to a angular velocity $\Omega=  0.071$ $\text{fm}^{-1}$ with $\Bar{\mu}=$ 0, 100 MeV.}
    \label{A2}
\end{figure}

Figure~\ref{Tau1} shows the relaxation time $\tau$ for quarks as a function of $T$. The calculation is performed for a collision energy $\sqrt{s_{NN}}=$ 200 GeV, which corresponds to an angular velocity $\Omega=$ 0.052 $\text{fm}^{-1}$ and chemical potentials $\mu=$ 0, 100 MeV. 

Figure~\ref{Tau2} shows the relaxation time $\tau$ for quarks as a function of $T$, this time for a collision energy $\sqrt{s_{NN}}=$ 10 GeV, which corresponds to a angular velocity $\Omega=$ 0.071 $\text{fm}^{-1}$ and two values of the chemical potential $\mu=$ 0, 100 MeV. In both cases, $\tau<$ 5 fm for the considered temperature range. 

Figures~\ref{A1} and \ref{A2} show the relaxation time $\Bar{\tau}$ for antiquarks as a function of $T$ obtained for a collision energy $\sqrt{s_{NN}}=$ 200 and 10 GeV, which correspond to a angular velocities $\Omega=$ 0.052 and 0.071 $\text{fm}^{-1}$, respectively, for chemical potentials $\mu=$ 0, 100 MeV. Notice that for the largest antiquark chemical potential the relaxation times are larger. This behavior is opposite to that of the quarks, where the relaxation time is lower for larger quark chemical potentials. However, also notice that $\Bar{\tau}<$ 6 fm for the considered temperature range.

\begin{figure}[t]
    \centering
    \includegraphics[scale=0.45]{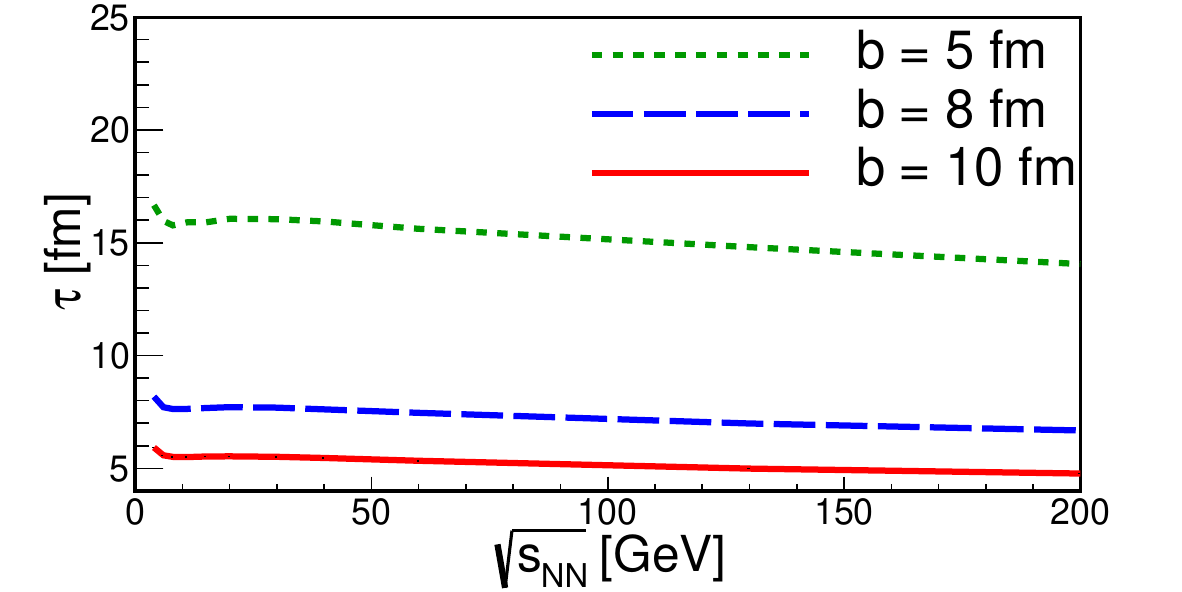}
    \includegraphics[scale=0.45]{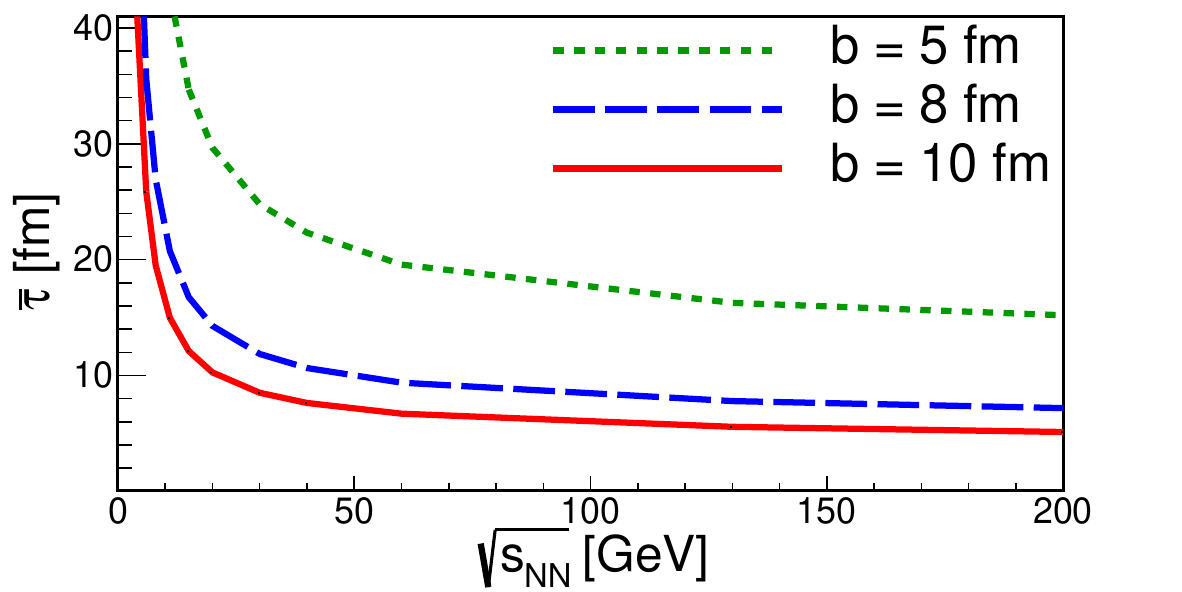}
    \caption{Top: relaxation time $\tau$ for quarks as a function of $\sqrt{s_{NN}}$ for semicentral  collisions at impact parameters $b$ = 5, 8 and 10 fm. Bottom: relaxation time $\bar{\tau}$ for antiquarks as a function of $\sqrt{s_{NN}}$ for semicentral  collisions at impact parameters $b$ = 5, 8 and 10 fm.}
    \label{TauEner}
\end{figure}

Figure~\ref{TauEner} shows the relaxation time for quarks (top) and antiquarks (bottom) as functions of $\sqrt{s_{NN}}$ for semicentral collisions at impact parameters $b=5$, 8 and 10 fm. For each value of $\sqrt{s_{NN}}$, the temperature $T$ and and maximum baryon chemical potential $\mu_B=3\mu$ at freeze-out were extracted from the parametrization in Ref.~\cite{cleymans2006comparison}
\begin{equation}
\begin{aligned}
%  T(\mu_B)&=166-139\mu_B^2-53\mu_B^4,\\
%  \mu_B(\sqrt{s_{NN}})&=\frac{1308}{1000+0.273\sqrt{s_{NN}}},
  T(\mu_B)&=0.166-0.139\mu_B^2-0.053\mu_B^4,\\
  \mu_B(\sqrt{s_{NN}})&=\frac{1.308}{1+0.273\sqrt{s_{NN}}},
\end{aligned}
\end{equation}
where $\mu_B$ and $\sqrt{s_{NN}}$ are given in GeV. Also, the values for $\Omega$ were obtained from the parametrization found in Ref.~\cite{Ayala:2020ndx}
\begin{equation}
    \Omega=\frac{b^2}{2V_N}\left[1+2\left( \frac{m_N}{\sqrt{s_{NN}}}\right)^{1/2} \right],
\end{equation}
where $V_N=\frac{4}{3}\pi R^3$. The relaxation times for quarks and antiquarks exhibit overall a decrease as functions of $\sqrt{s_{NN}}$. Notice that the relaxation times are smaller for the quark case than for the antiquark case. Also, for the largest impact parameters considered, the relaxation times for the quark case in the energy range considered are smaller than 10 fm, which is the ballpark lifetime of the QGP in heavy-ion reactions. However, for the antiquark case, this is true only for energies $\sqrt{s_{NN}}\gtrsim 50$ GeV. This indicates that, although quarks are likely to align their spins within the lifetime of the QGP, this is not the case for the antiquarks at least for small collision energies.

\begin{figure}[t]
    \centering
    \includegraphics[scale=0.45]{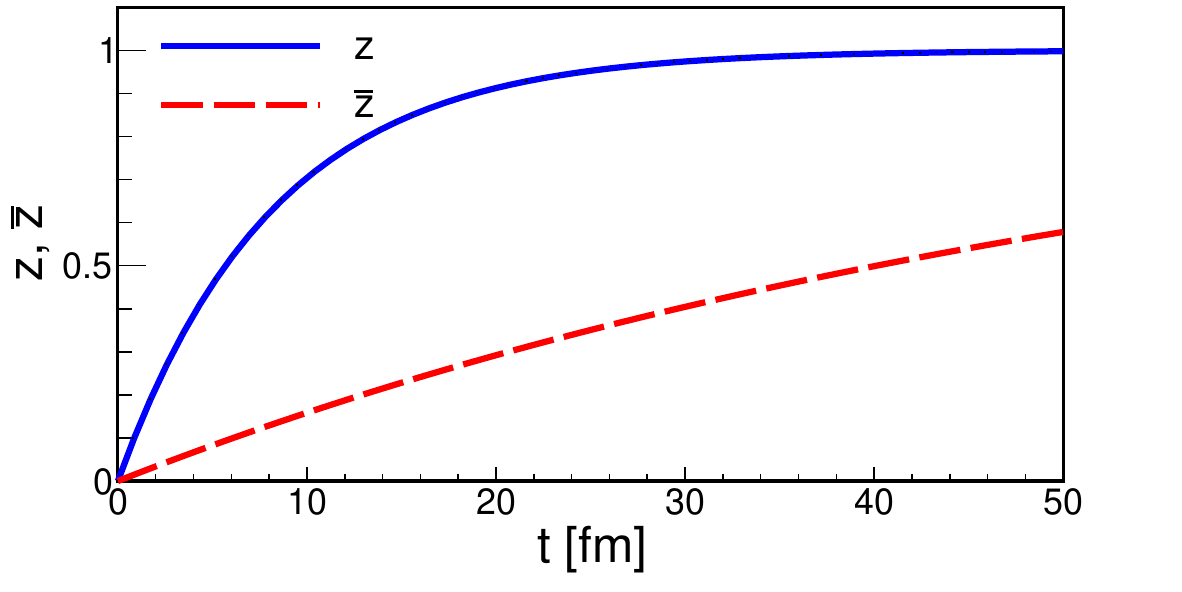}
    \caption{Intrinsic polarization for quarks ($z$) and antiquarks ($\bar{z}$) as functions of time $t$ for semicentral  collisions at an impact parameter $b=10$ fm for $\sqrt{s_{NN}}=4$ GeV. }
    \label{Pol}
\end{figure}

Recall that the relaxation time can be used to define the {\it intrinsic polarization} as the probability to polarize the quark spin along the direction of the angular velocity as a function of time. When an initial number of particles $N_0$, originally unpolarized, is placed in the rotating medium, the number of particles that remain unpolarized varies as a function of time $t$ as $N=N_0\exp(-t/\tau)$. Therefore, the number of particles in the polarized state is given by $N=N_0\left[1-\exp(-t/\tau)\right]$. The factor $\left[1-\exp(-t/\tau)\right]$ is therefore the intrinsic polarization. Figure~\ref{Pol} shows the intrinsic polarization for
quarks ($z$) and antiquarks ($\bar{z}$), given by
\begin{equation}
    \begin{aligned}
        z&=1-e^{-t/\tau},\\
        \bar{z}&=1-e^{-t/\bar{\tau}},
    \end{aligned}
\end{equation}
as functions of time $t$ for semicentral collisions at an impact parameter $b=8$ fm and a collision energy $\sqrt{s_{NN}}=4$ GeV. Notice that $z$ approaches 1 faster that $\bar{z}$.

\section{Summary and conclusions}\label{concl}

In this work, we used a thermal field theoretical framework to compute the relaxation times for massive quarks and antiquarks to align their spins with the angular velocity in a rigidly rotating medium at finite temperature and baryon density. The rigid rotation is implemented using the recently found fermion propagator immersed in a cylindrical rotating environment. In principle, the effects of rotation could also be included into the properties of the gluon propagator. However, notice that the kinematical gluon momentum region that contributes to the calculation corresponds to Landau damping and thus to soft modes. The main role of these modes at finite temperature and
baryon density is to mediate the interaction between plasma quarks and the test quark whose spin alignment with the angular velocity has been monitored. Notice that the energy associated to a typical angular velocity for semicentral
collisions is of order $\Omega\sim 0.05$ $\text{fm}^{-1}\sim 10$ MeV. In this
sense, including the effects of rotation into the gluon propagator with a temperature of order $T\sim100$ MeV, although not negligible, represents a subleading effect of order $10\%$.\\
The relaxation time is computed as the inverse of the interaction rate to produce an asymmetry between quark (antiquark) spin projections pointing along and opposite to the angular velocity. We found that for conditions resembling a heavy-ion collision the relaxation times for quarks are within the putative life-time of the QGP. However, for antiquarks this is the case only for collision energies $\sqrt{s_{NN}}\gtrsim 50$ GeV. We quantified these results in terms of the intrinsic quark and antiquark polarizations, that is, the probability to build the spin asymmetry as a function of time. Our results show that these intrinsic polarizations tend to 1 with time at different rates given by the relaxation times $\tau$ and $\Bar{\tau}$ with quarks building the asymmetry at a faster pace. These intrinsic polarizations are essential ingredients to describe the polarization of $\Lambda$ and $\overline{\Lambda}$ hyperons in relativistic heavy-ion collisions. The consequences of the results hereby found are currently being explored and will be reported elsewhere.
\\

\section*{Acknowledgements}

Support for this work was received in part by UNAM-PAPIIT-IG100322 and by Consejo Nacional de Humanidades, Ciencia y Tecnolog\'ia grant numbers CF-2023-G-433, A1-S-7655 and A1-S-16215. S.B.-L. and J.J.M.-S. acknowledge the financial support of corresponding fellowships granted by Consejo Nacional de Humanidades, Ciencia y Tecnología as part of the Sistema Nacional de Posgrados.
% 

%\subfile{formulas.tex}

%\begin{acknowledgements}
%
%\end{acknowledgements}

%\bibliographystyle{spphys}

\bibliography{biblio}

\end{document}